\documentclass[11pt]{article}
\begin{document}
\title{\bf {Is Minisuperspace Quantum Gravity Reliable?}}
\author{O. Saremi\footnote{Omid@physics.sharif.ac.ir}}
\date{\small{Feb.12 , 2001}}
\maketitle
\centerline{\textit{Department of Physics}}
\centerline{\textit{Sharif University of Technology,Azadi St.}}
\centerline{\textit{Tehran, Iran P.O. Box: 11365-9161} }

\begin{abstract}
We study a minisuperspace quantum cosmology for a 2+1 dimensional
de Sitter universe and find the wave function both exactly and in
WKB approximation. Then we extend the model to a canonically
quantized field theory for quantum gravity, i.e., a
midisuperspace, and obtain the wave functional of the resulting
field theory in the saddle point approximation. It is shown that
these two approaches yield different results.
\end{abstract}

\section{Introduction}

Quantum cosmology (QC) was initiated by B.S.Dewitt in his seminal
paper on Hamiltonian quantization of gravity [1]. Up to advent of
reasonable boundary conditions to pick out a unique solution to
the Wheeler-Dewitt equation (WDW), QC was out of attention. In
the early eighties, proposals on possible boundary conditions
revitalized QC. [2,3,4].
Equivalent to Hamiltonian quantization of 4-dimensional gravity, one can path integrate over all metric configurations and sum over all possible 4-topologies which have a given $%
\partial M$ as their boundary [5]. The space of all configurations for
4-metric is called superspace (SS) where diffeomorphically
equivalent configurations are factored out. Mathematically,
dealing with the full SS is too difficult, if not impossible.
Inevitably, we must invoke to some approximate schemes. A
commonly used approximate model is called "minisuperspace" model
[1]. In this approximation, one confines his attention to a
restricted region of SS. In practice, one freezes or suspends many
infinite degrees of freedom of gravitational field on a time
constant slice of 3-geometry and retains a few of them alive.

Despite the problems surrounding the interpretation of the wave
function of the universe, by using some generally accepted
interpretational rules [4], many attempts have been made to
present explanations for some observed features of the universe
in the context of "minisuperspace" models, e.g., cosmological
constant problem [7] or lacking the existence of primordial black
holes at the present state of the universe [8]. There are some
other and more realistic approaches. Inhomogenous perturbations
and midisuperspace models are examples of more realistic
approximations to the full quantized theory. Inhomogeneous
perturbations approach to quantum cosmology is a perturbative way
to pass beyond the minisuperspace models [9]. In a typical
midisuperspace model, one retains some dependence of metric
configurations on spatial coordinates. This means that the less
symmetric configurations are considered as the space of path
integration. The spherically symmetric midisuperspaces, the BCMN
models [10], have been used as models for quantized Schwarzchild
black holes. In the light of these models some of expected
features of quantum black holes have been derived successfully
[11]. Also a Hamiltonian quantization of false-vacuum bubbles
within the framework of a spherically symmetric midisuperspace
has been presented [12]. There is no support in favor of this
claim that the minisuperspace configurations have the dominant
contribution to the partition function of quantum gravity.
Moreover minisuperspace is not known to be a part of a systematic
approximation to the full theory. Therefore, it is possible that
some of derived predictions from these minisuperspace
calculations to be artifact of this approximation. In this way, a
comparison between amplitudes derived from a minisuperspace
calculation and of midi one, will be useful.

 Through this paper, it is shown that a minisuperspace wave function can differ from midi one. The organization of the
paper is as follows:

 In section 2, we take a minisuperspace for QC in 2+1 dimensions and obtain the ''wave function'' in saddle point approximation. Through Section 3, we will give the exact solutions to the Wheeler-Dewitt equation. Section 4 has been devoted to extension of
this minisuperspace model to a midi one. We solve the resulting
field theory in the saddle point approximation and obtain the wave
functional of the theory.

Finally, we present a comparison between the obtained wave
functions and wave functionals.
 Paper will end up with a conclusion.

\section{Minisuperspace Saddle Point Wave Functions }

In this section, we review the quantum cosmology of a 2+1
dimensional de Sitter universe.  Take the following 2+1
dimensional FRW line element as a minisuperspace:

\begin{equation}
 ds^{2}=-N(t)^{2}+a(t)^{2}(\frac{dr^{2}}{1-r^{2}}+r^{2}d\theta^{2}).
\end{equation}
 Spatial section of the 3-manifold taken to be $S^{2}$, i.e.,
k=1, to avoid getting infinite action due to infinite extension
of a spatially flat k=0 or hyperbolic k=-1 universe. Our
minisuperspace obviously contains classical de Sitter (dS3)
universes.

 Substituting the above ansatz into the Einstein-Hilbert action leads us to:
\begin{equation}
S=-\frac{m_{p}^{2}}{4\pi }\int
(\frac{\stackrel{.}{a}^{2}}{N^{2}}-N+\Lambda Na^{2})dt.
\end{equation}
Varying with respect to $a(t)$ and N yields an equation of motion
\begin{equation}
\frac{d}{dt}(\frac{\stackrel{.}{a}}{N})=\Lambda Na,
\end{equation}
 and the Hamiltonian constraint:
\begin{equation}
H=\frac{1}{4}\pi _{a}^{2}+1-\Lambda a^{2}=0,
\end{equation}

where $\pi _a$ is momentum conjugate of $a$. There is a solution
to equations (3) and (4) for $\Lambda >0:$
\begin{equation}
a(t)=\frac{1}{\sqrt{\Lambda }}\cosh (\sqrt{\Lambda }t+\beta ),
\end{equation}
where $\beta $ is an integration constant. This solution describes
classical dynamics of a dS3 universe.

Euclideanized version of the action, resulting equation of motion
and the Hamiltonian constraint read:
\begin{equation}
S_{E}=-\frac{m_{p}^{2}}{4\pi }\int (\frac{\stackrel{.}{a}^{2}}{N^{2}}%
+N-\Lambda Na^{2})d\tau,
\end{equation}
\begin{equation}
\frac{d}{d\tau }(\frac{\stackrel{.}a}{N})=-\Lambda Na,
\end{equation}
\begin{equation}
H_{E}=\frac{1}{4}\pi _{a}^{2}-1+\Lambda a^{2}=0.
\end{equation}
A solution to these set of equations which meets the requirements
of a no boundary (NB) instanton [8] is:
\begin{equation}
a(\tau )=\frac{1}{\sqrt{\Lambda }}\sin (\sqrt{\Lambda }\tau ).
\end{equation}
Note that the condition $\frac{da}{dt}=1$, necessary for regular
closing off of four geometry at singularity a(0)=0, is
automatically satisfied by the Hamiltonian constraint. For
a$_{\partial M}<\frac{1}{\sqrt{\Lambda }}$, there is a ''real''
instanton which is a portion of a $S^{3}$ sphere [8].

For a$_{\partial M}>\frac{1}{\sqrt{\Lambda }}$, In complex plane
of $\tau $, we should choose a path along $\tau _{Re}$ axis to
$\tau _{\max }=\frac{\pi }{2\sqrt{\Lambda }}$ which determines
the maximum radius of such a compact instanton. This part of
instanton describes one half of a $S^{3}$ sphere. Choosing the
path to continue parallel to the $\tau _{Im}$ axis to a given $a
>\frac{1}{\sqrt{\Lambda}}$ on the boundary, a($\tau $) still
remains real:
\begin{equation}
a(\frac{\pi }{2\sqrt{\Lambda }}+i\tau _{Im}) =\frac{1}{\sqrt{\Lambda }}%
\cosh (\sqrt{\Lambda }\tau _{Im}).
\end{equation}

This part of the instanton describes half of a Lorentzian dS3
universe. There is also another instanton which satisfies NBP
conditions and contributes to the saddle point approximation [6].

Resulting NB wave function, for $a<\frac{1}{\sqrt{\Lambda }}$ will
be:
\begin{equation}
\psi _{NB}\sim\exp{(-2a\sqrt{1-\Lambda
a^{2}}-\frac{2}{\sqrt{\Lambda }}\sin ^{-1}\sqrt{\Lambda }a)}
\end{equation}

, and for $a >\frac{1}{\sqrt{\Lambda }}$:
\begin{equation}
\psi _{NB}\sim e^{-\frac{\pi }{\sqrt{\Lambda }}}\cos
(2a\sqrt{1-\Lambda
a^{2}}-\frac{2}{\sqrt{\Lambda }}\cosh ^{-1}\sqrt{\Lambda }a-\frac{\pi }{4}%
).
\end{equation}

Vilenkin wave function in WKB approximation [3,4] will have the
following form for $a<\frac{1}{\sqrt{\Lambda }}$:
\begin{equation}
\psi _{V}\sim\exp{(2a\sqrt{1-\Lambda
a^{2}}+\frac{2}{\sqrt{\Lambda }}\sin ^{-1}\sqrt{\Lambda }a)}
\end{equation}
, and for $a >\frac{1}{\sqrt{\Lambda }}$:
\begin{equation}
\psi _{V}\sim \exp{(\frac{\pi }{\sqrt{\Lambda
}})}\exp{[-i(2a\sqrt{1-\Lambda a^{2} }-\frac{2}{\sqrt{\Lambda
}}\cosh ^{-1}\sqrt{\Lambda }a)]}.
\end{equation}
\section{Exact Solutions}

According to Dirac prescription for quantization [13], the wave
function of a constraint system should be annihilated by the
operator version of classical constraints. Replacing $\pi _{a}$ by $-i\hbar \frac{\partial }{%
\partial a}$ in the Hamiltonian constraint, we will find the following
schr\"{o}dinger like equation, a Wheeler-Dewitt equation [1], for
the wave function of a dS3 universe:
\begin{equation}
-\frac{1}{4a^{P}}\frac{\partial }{\partial a}(a^{P}\frac{\partial \psi }{%
\partial a})+(1-\Lambda a^{2})\psi =0,
\end{equation}
 where P carries some part of factor ordering
ambiguity due to indefiniteness of measure of path integral or
equivalently quadratic form of Hamiltonian in $\pi _{a}$. We will
set it to 0.

There are exact solutions to the differential equation(15) in
terms of Whittaker functions of type W and M [14]:
\begin{equation}
\psi (a,\Lambda )=\frac{\zeta _{m}}{\sqrt{a}}WM(-\frac{1}{4\sqrt{-\Lambda }},%
\frac{1}{4},\sqrt{-\Lambda }a^{2})+\frac{\zeta _{w}}{\sqrt{a}}WW(-\frac{1}{4%
\sqrt{-\Lambda }},\frac{1}{4},\sqrt{-\Lambda }a^{2}).
\end{equation}
Constant coefficients $\zeta _m$ and $\zeta _w$ should be
determined by considering appropriate boundary conditions.

Asymptotic expansion of $\psi (a,\Lambda )$ can be obtained
easily [9]:
\begin{eqnarray}
&&\psi (a,\Lambda )\sim \frac{1}{\sqrt{a}}\Lambda ^{\frac{i}{8\sqrt{\Lambda }%
}}(\frac{(i-1)\sqrt{2\pi }}{4\Gamma (\frac{3}{4}+\frac{i}{4\sqrt{\Lambda }})%
}e^{\frac{1}{4\sqrt{\Lambda }}}\zeta _{m}+\zeta _{w})a^{\frac{i}{2\sqrt{%
\Lambda }}}e^{-\frac{i}{2}\sqrt{\Lambda}a^{2}}\nonumber\\
&&+\frac{1}{\sqrt{a}}(\Lambda^{-\frac{i}{8\sqrt{\Lambda }}}\frac{\sqrt{\pi }}{2\Gamma (\frac{3}{4}-\frac{%
i}{4\sqrt{\Lambda }})})\zeta _{m}a^{-\frac{i}{2\sqrt{\Lambda }}}e^{\frac{i%
}{2}\sqrt{\Lambda }a^{2}}.
\end{eqnarray}
 $a\rightarrow \infty $

Vilenkin wave function, has only outgoing sector at very large
values of $a$, therefore:
\begin{equation}
\zeta _{w}=-\frac{(i-1)\sqrt{2\pi }}{4\Gamma (\frac{3}{4}+\frac{i}{4\sqrt{%
\Lambda }})}e^{\frac{1}{4\sqrt{\Lambda }}}\zeta _{m}.
\end{equation}
We obtain the Vilenkin wave function up to a constant $\gamma $:
\begin{equation}
\psi _{V}(a,\Lambda )=\gamma \{\frac{1}{\sqrt{a}}WM(-\frac{1}{4\sqrt{%
-\Lambda }},\frac{1}{4},\sqrt{-\Lambda }a^{2})-\frac{1}{\sqrt{a}}\frac{(i-1)%
\sqrt{2\pi }}{4\Gamma (\frac{3}{4}+\frac{i}{4\sqrt{\Lambda }})}e^{\frac{1}{4%
\sqrt{\Lambda }}}WW(-\frac{1}{4\sqrt{-\Lambda }},\frac{1}{4},\sqrt{-\Lambda }%
a^{2})\}.
\end{equation}
To find the Hartle-Hawking state, we note that the ingoing and
outgoing sectors of $\psi (a,\Lambda )$ should have the same
amplitudes for $a\rightarrow \infty .$ This implies:
\begin{equation}
\zeta _{w}=(\frac{\sqrt{\pi }}{2\Gamma (\frac{3}{4}-\frac{i}{4\sqrt{%
\Lambda }})}-\frac{\sqrt{2\pi }(i-1)}{4\Gamma (\frac{3}{4}+\frac{i}{4\sqrt{%
\Lambda }})}e^{\frac{1}{4\sqrt{\Lambda }}})\zeta _{m}.
\end{equation}
Therefore, the resulting Hartle-Hawking state will be:
\begin{eqnarray}
&&\psi _{H-H}(a,\Lambda )=\frac{\zeta _{m}}{\sqrt{a}}WM(-\frac{1}{4\sqrt{%
-\Lambda }},\frac{1}{4},\sqrt{-\Lambda }a^{2})\nonumber\\
&&+\frac{1}{\sqrt{a}}(\frac{\sqrt{%
\pi }}{2\Gamma (\frac{3}{4}-\frac{i}{4\sqrt{\Lambda }})}-\frac{\sqrt{2\pi }%
(i-1)}{4\Gamma (\frac{3}{4}+\frac{i}{4\sqrt{\Lambda }})}e^{\frac{1}{4\sqrt{%
\Lambda }}})\zeta _{m}WW(-\frac{1}{4\sqrt{-\Lambda }},\frac{1}{4},\sqrt{%
-\Lambda }a^{2}).
\end{eqnarray}
\section{ Midisuperspace Saddle Point Wave Functionals}

Is there any dramatic discrepancy between using a minisuperspace
instead of full theory or even a midisuperspace? A way to answer
to this question is to compare a minisuperspace amplitude with a
midi one. Here we present and, approximately, solve a ''midi''
superspace with axial symmetry for quantum cosmology of dS3
universes which is described by the following line element:
\begin{equation}
ds^{2}=-N(t,r)^{2}dt^{2}+\phi (t,r)dr^{2}+\psi (t,r)d\theta ^{2}.
\end{equation}
By axial symmetry we mean that the metric components are considered to be $%
\theta $ independent.

Substituting the above metric ansatz, Lagrangian density reads:
\begin{equation}
\pounds =N(\psi \phi )^{\frac{1}{2}}[\frac{1}{2N^{2}\phi \psi }(-\stackrel{.%
}{\phi }+2N_{r}^{^{\prime }}-\frac{\phi ^{^{\prime }}}{\phi }N_{r})(-%
\stackrel{.}{\psi }+\frac{\psi ^{^{\prime }}}{\phi
}N_{r})+^{2}R+2\Lambda]
\end{equation}

, where $^{2}R$ is the scalar curvature of spatial sector of three geometry .$%
^{2}R$ is given by:
\begin{equation}
^{2}R=\frac{2\psi ^{^{\prime \prime }}\psi \phi -\psi ^{^{\prime
2}}\phi -\phi ^{^{\prime }}\psi \psi ^{^{\prime }}}{2\psi
^{2}\phi ^{2}}.
\end{equation}
Momentum conjugates to the field variables are simply:
\begin{equation}
\pi _{\phi }=\frac{\partial \pounds }{\partial \stackrel{.}{\phi }}=\frac{1%
}{2N(\phi \psi )^{\frac{1}{2}}}(\stackrel{.}{\psi }-\frac{\psi ^{^{\prime }}%
}{\phi }N_{r}),
\end{equation}
\begin{equation}
\pi _{\psi }=\frac{\partial \pounds }{\partial \stackrel{.}{\psi }}=\frac{1%
}{2N(\phi \psi )^{\frac{1}{2}}}(\stackrel{.}{\phi }+\frac{\phi ^{^{\prime }}%
}{\phi }N_{r}-2N_{r}^{^{\prime }}).
\end{equation}
A Legender transformation will result in the Hamiltonian
constraint density:
\begin{equation}
H=\frac{1}{2N(\phi \psi )^{\frac{1}{2}}}[ \stackrel{.}{\phi } (%
\stackrel{.}{\psi }-\frac{\psi ^{^{\prime }}}{\phi }N_{r})+ \stackrel{.}{%
\psi }(\stackrel{.}{\phi }+\frac{\phi ^{^{\prime }}}{\phi }%
N_{r}-2N_{r}^{^{\prime }})]-\pounds,
\end{equation}
\begin{equation}
H=\frac{1}{2N(\phi \psi )^{\frac{1}{2}}}(\stackrel{.}{\psi }\stackrel{.}{%
\phi }+N_{r}^{2}\frac{\psi ^{^{\prime }}\phi ^{^{\prime }}}{\phi ^{2}}-2%
\frac{N_{r}N_{r}^{^{\prime }}}{\phi }\psi ^{^{\prime
}})-N\frac{2\psi ^{^{\prime \prime }}\psi \phi -\psi ^{^{\prime
2}}\phi -\phi ^{^{\prime }}\psi \psi ^{^{\prime }}}{2\psi
^{\frac{3}{2}}\phi
^{\frac{3}{2}}}-2N(\phi\psi)^{\frac{1}{2}}\Lambda.
\end{equation}
Now, we should express the time derivatives of the filed variables
in terms of momenta, fields and their spatial derivatives:
\begin{equation}
\stackrel{.}{\phi }=2N(\phi \psi )^{\frac{1}{2}}\pi _{\psi
}+2N_{r}^{^{\prime }}-\frac{\phi ^{^{\prime }}}{\phi }N_{r},
\end{equation}
\begin{equation}
\stackrel{.}{\psi }=2N(\phi \psi )^{\frac{1}{2}}\pi _{\phi
}+\frac{\psi ^{^{\prime }}}{\phi }N_{r}.
\end{equation}
Integration over a spacelike hypersurface $\Sigma $ leads to the
Hamiltonian constraint:
\begin{equation}
H=2\pi \int_{\Sigma }\{N[2(\phi \psi )^{\frac{1}{2}}\pi _{\phi
}\pi _{\psi
}-(\phi \psi )^{\frac{1}{2}2}R-2\Lambda (\phi \psi )^{\frac{1}{2}}]+N_{r}(%
\frac{\psi ^{^{\prime }}}{\phi }\pi _{\psi }-\frac{\phi ^{^{\prime }}}{\phi }%
\pi _{\phi }-2\pi _{\phi }^{^{\prime }})\}dr.
\end{equation}
Variation with respect to $N$ and $N_{r}$ leads into the
Hamiltonian and the  momentum constraints:
\begin{equation}
H_{0}=2(\phi \psi )^{\frac{1}{2}}\pi _{\phi }\pi _{\psi }-2\Lambda
(\phi \psi )^{\frac{1}{2}}-\frac{2\psi ^{^{\prime \prime }}\psi
\phi -\psi
^{^{\prime 2}}\phi -\phi ^{^{\prime }}\psi \psi ^{^{\prime }}}{2\psi ^{\frac{%
3}{2}}\phi ^{\frac{3}{2}}}=0 ,
\end{equation}
\begin{equation}
H_{r}=\frac{\psi ^{^{\prime }}}{\phi }\pi _{\psi }-\frac{\phi ^{^{\prime }}%
}{\phi }\pi _{\phi }-2\pi _{\phi }^{^{\prime }}=0.
\end{equation}
Hamiltonian constraint carries time reparametrization invariance
of the classical theory and momentum constraint is the generator
of infinitesimal coordinate transformations within spacelike
hypersurfaces.

The number of field variables and constraints are equal, then it
is tempting to solve $\pi _{\phi }$ and $\pi _{\psi }$ in terms of
field variables and their spatial derivatives by constructing a
superposition of constraints to eliminate $\pi _{\psi }.$

Note that $^{2}R$ can be rewritten as follows:
\begin{equation}
^{2}R=\frac{1}{2\psi ^{^{\prime }}}\partial _{r}(\frac{\psi ^{^{\prime }2}}{%
\psi \phi }).
\end{equation}
Now, consider the below superposition:
\begin{equation}
2(\phi \psi )^{\frac{1}{2}}\frac{\phi }{\psi ^{^{\prime }}}\pi
_{\phi }H_{r}-H_{0}=\frac{(\phi \psi )^{\frac{1}{2}}}{\psi
^{^{\prime }}}(-2\phi ^{^{\prime }}\pi _{\phi }^{2}-4\phi \pi
_{\phi }\pi _{\phi }^{^{\prime }}+^{2}R\psi ^{^{\prime }}+2\Lambda
\psi ^{^{\prime }})=0.
\end{equation}
By considering the eqs.(34) and (35), we obtain:
\begin{equation}
\partial _{r}(-2\phi \pi _{\phi }^{2}+\frac{\psi ^{^{\prime
}2}}{2\psi \phi }+2\Lambda \psi )=0
\end{equation}
, which simply results in:
\begin{equation}
\pi _{\phi }=\sqrt{\frac{1}{2\phi }(\frac{\psi ^{^{\prime 2}}}{2\psi \phi }%
+2\Lambda \psi -2)}.
\end{equation}
Regarding eq.(32), $\pi _{\psi }$ will become:
\begin{equation}
\pi _{\psi }=\frac{\frac{1}{2\psi ^{^{\prime }}}(\frac{\psi ^{^{\prime }2}}{%
\psi \phi })^{^{\prime }}+2\Lambda }{2\sqrt{\frac{1}{2\phi
}(\frac{\psi ^{^{\prime 2}}}{2\psi \phi }+2\Lambda \psi -2)}}.
\end{equation}
The leading term in a semiclassical expansion of the \ ''wave
functional'' of this field theory will be $e^{ \pm {iS_{cl}}}$,
where $S_{cl}$ is the action evaluated for classical path:
\begin{equation}
\delta S=\int dr(\pi _{\phi }\delta \phi +\pi _{\psi }\delta \psi
).
\end{equation}
We can choose a specific contour of integration to simplify
integration process. First of all, we hold $\psi $ constant and
then integrate over $\phi $, to a configuration of $\phi $ such
that:
\begin{equation}
\frac{\psi ^{^{\prime 2}}}{2\psi \phi }+2\Lambda \psi -2=0.
\end{equation}
This configurations are the ''boundary'' between classically
allowed and forbidden regions. After that, we will hold $\phi $
constant such that eq.(40) holds and integrate over $\psi $ to a
given configuration.

Before going ahead, we note that:
\begin{equation}
\pi _{\psi }=\frac{\frac{1}{2\psi ^{^{\prime }}}(\frac{\psi ^{^{\prime }2}}{%
\psi \phi })^{^{\prime }}+2\Lambda }{2\sqrt{\frac{1}{2\phi
}(\frac{\psi ^{^{\prime 2}}}{2\psi \phi }+2\Lambda \psi
-2)}}=\frac{(\sqrt{\frac{\psi ^{^{\prime 2}}}{2\psi \phi
}+2\Lambda \psi -2})^{^{\prime }}}{\psi ^{^{\prime
}}\sqrt{\frac{1}{2\phi }}},
\end{equation}
then this part will not contribute to the action. Therefore, the
action reads:
\begin{equation}
S=\int \sqrt{\frac{1}{2\phi }(\frac{\psi ^{^{\prime 2}}}{2\psi
\phi }+2\Lambda \psi -2)}d[\phi]dr.
\end{equation}
Integration is straightforward and results in :
\begin{equation}
S=\int(2\sqrt{\frac{\psi ^{^{\prime }2}}{4\psi }-(1-\Lambda \psi )\phi }-2\sqrt{%
\frac{\psi ^{^{\prime }2}}{4\psi }}Arctanh(\frac{\sqrt{\frac{\psi
^{^{\prime }2}}{4\psi }-(1-\Lambda \psi )\phi }}{\sqrt{\frac{\psi
^{^{\prime }2}}{4\psi }}}))dr.
\end{equation}
Wave functional for classically forbidden region will be a
superposition of exponentially decaying and growing forms and in
the allowed region is a linear superposition of oscillating
exponentials of the action.

Now, we are in the position to calculate the wave functional of a
2+1 dimensional FRW-like universe. It will be sufficient to
consider a homogeneous and isotropic universe on the final
hyprsuface of simultaneity.

By substituting ''FRW'' form for the metric on the final
hypersurface and regarding to eq.(43), we will reach the
following saddle point ''wave functionals'':

Vilenkin wave functional in tunneling region will be:
\begin{equation}
\Psi _{V}\sim e^{2a\sqrt{1-\Lambda a^{2}}-2a\int_{0}^{1}\tan ^{-1}[\frac{r}{%
(1-r^{2})^{\frac{1}{2}}}\sqrt{1-\Lambda a^{2}}]dr}
\end{equation}
, and in classical region:
\begin{equation}
\Psi _{V}\sim e^{-i(2a\sqrt{-1+\Lambda a^{2}}-2a\int_{0}^{1}\tan ^{-1}[%
\frac{r}{(1-r^{2})^{\frac{1}{2}}}\sqrt{-1+\Lambda a^{2}}]dr}.
\end{equation}
Hartle-Hawking wave functional in tunneling region will be:
\begin{equation}
\Psi _{H-H}\sim e^{^{-2a\sqrt{1-\Lambda a^{2}}+2a\int_{0}^{1}\tan ^{-1}[%
\frac{r}{(1-r^{2})^{\frac{1}{2}}}\sqrt{1-\Lambda a^{2}}]dr}}
\end{equation}
, and in classical region:
\begin{equation}
\Psi _{H-H}\sim \cos (2a\sqrt{-1+\Lambda a^{2}}-2a\int_{0}^{1}\tanh ^{-1}[%
\frac{r}{(1-r^{2})^{\frac{1}{2}}}\sqrt{-1+\Lambda
a^{2}}]dr-\frac{\pi }{4}).
\end{equation}

As obviously can be seen from the resulting midisuperspace wave
functionals, there is a discrepancy between minisuperspace wave
functions and midisuperspace wave functionals.
\section{Conclusion} Minisuperspace approximation to find the wave
function of a 2+1 dimensional de Sitter universe, seems to be
unreliable and a more realistic model such as a midisuperspace
yields different result. Therefore, certain predictions derived
from minisuperspace models may be wrong. Comparing predictions
based on a canonically quantized field theory for quantum gravity
(a midisuperspace) with minisuperspace one, e.g., tunneling
amplitudes for creation of a dS3 universe or isotropy of a large
universe are subjects of further investigations.
\section{Acknowledgement}

I am deeply grateful to Prof.R.Mansouri for his illuminating
directions, Prof.S.Rouhani for very inspiring discussions and
A.T.Rezakhani for his kindly helps in providing the LaTex version
of the paper.

\section{References}

\hspace{4mm} 1. B.S.Dewitt, Phys.Rev.160(1967)1113.

2. J.B.Hartle and S.W.Hawking, Phys.Rev.D28, 2960(1983).

3. A.Vilenkin, Phys.Rev.D33, 3560(1986).

4. A.Vilenkin, Phys.Rev.D37, 888(1988).

5. G.W.Gibboins, S.W.Hawking, Euclidean Quantum Gravity, World
Scientific(1993)

6. P.D.D'eath, Supersymmetric Quantum Cosmology, Cambridge
University Press(1995).

7. A.Strominger, Nucl.Phys.B319(1989) 722-732

8. R.Bousso and S.W.Hawking, Phys.Rev.D52, No.10, 5659(1995)

9. J.J.Halliwell, S.W.Hawking, Phys.Rev.D31, No.8, 1777(1985).

10. B.K.Berger, D.M.Chitre, V.E.Moncrief, and Y.Nutku,
Phys.Rev.D5, 2467(1972).

11. C.Vaz, Phys.Rev.D61(2000), 064017.

12. W.Fischler,D.Morgan, and J.Polchinski, Phys.Rev.D42, No.12
(1990).

13. P.A.M.Dirac, Lectures on quantum mechanics (New York, Belfer
Graduate School of Science, Yeshiva University, 1964).

14. M.Abramowitz and I.Stegun, Hand Book of Mathematical
Functions, Dover Publication, INC., New York, 1965.

\end{document}